\documentclass[letterpaper]{article} %
\usepackage{aaai24}
\usepackage{times}
\usepackage{helvet}
\usepackage{courier}
\usepackage[hyphens]{url}
\usepackage{graphicx}
\usepackage{natbib}
\usepackage{caption}
\usepackage{subcaption}
\usepackage{booktabs}     
\usepackage{multirow}
\usepackage{mathtools,amssymb}
\usepackage{amsmath}
\usepackage{amsthm}
\usepackage{amsfonts}      
\usepackage{nicefrac}       
\usepackage{microtype}     
\usepackage{xcolor}         
\usepackage{algorithm,algorithmicx,algpseudocode}
\usepackage{bm}
\usepackage{graphicx}
\usepackage{enumitem}

\newcommand{\grad}{\nabla}

\newcommand{\bI}{\mathbf{I}}

\newcommand{\bzero}{\mathbf{0}}

\newcommand{\bx}{\mathbf{x}}

\newcommand{\bz}{\mathbf{z}}

\newcommand{\bepsilon}{{\boldsymbol{\epsilon}}}

\newcommand{\x}{{\boldsymbol x}}

\newcommand{\Ib}{{\boldsymbol I}}

\newcommand{\Nc}{{\mathcal N}}

\usepackage{comment}
\newcommand{\specialcell}[2][c]{%
\begin{tabular}[#1]{@{}c@{}}#2\end{tabular}}
\newcommand{\specialcellleft}[2][l]{%
\begin{tabular}[#1]{@{}l@{}}#2\end{tabular}}

\usepackage{newfloat}
\usepackage{listings}
\DeclareCaptionStyle{ruled}{labelfont=normalfont,labelsep=colon,strut=off} 
\floatstyle{ruled}
\newfloat{listing}{tb}{lst}{}
\floatname{listing}{Listing}

\title{Region-Disentangled Diffusion Model for High-Fidelity PPG-to-ECG Translation}
\author{
    Debaditya Shome\textsuperscript{\rm 1},
    Pritam Sarkar\textsuperscript{\rm 1, \rm 2},
    Ali Etemad\textsuperscript{\rm 1}
}
\affiliations{
    \textsuperscript{\rm 1}
    Queen's University, Canada\\
    \textsuperscript{\rm 2} Vector Institute\\
    \{debaditya.shome, pritam.sarkar, ali.etemad\}@queensu.ca 
}

\usepackage{bibentry}

\begin{document}

\maketitle

\begin{abstract}
The high prevalence of cardiovascular diseases (CVDs) calls for accessible and cost-effective continuous cardiac monitoring tools. Despite Electrocardiography (ECG) being the gold standard, continuous monitoring remains a challenge, leading to the exploration of Photoplethysmography (PPG), a promising but more basic alternative available in consumer wearables. This notion has recently spurred interest in translating PPG to ECG signals. In this work, we introduce Region-Disentangled Diffusion Model (RDDM), a novel diffusion model designed to capture the complex temporal dynamics of ECG. Traditional Diffusion models like Denoising Diffusion Probabilistic Models (DDPM) face challenges in capturing such nuances due to the indiscriminate noise addition process across the entire signal. Our proposed RDDM overcomes such limitations by incorporating a novel forward process that selectively adds noise to specific regions of interest (ROI) such as QRS complex in ECG signals, and a reverse process that disentangles the denoising of ROI and non-ROI regions. 
Quantitative experiments demonstrate that RDDM can generate high-fidelity ECG from PPG in as few as 10 diffusion steps, making it highly effective and computationally efficient. 
Additionally, to rigorously validate the usefulness of the generated ECG signals, we introduce CardioBench, a comprehensive evaluation benchmark for a variety of cardiac-related tasks including heart rate and blood pressure estimation, stress classification,  and the detection of atrial fibrillation and diabetes. Our thorough experiments show that RDDM achieves state-of-the-art performance on CardioBench. To the best of our knowledge, RDDM is the first diffusion model for cross-modal signal-to-signal translation in the bio-signal domain.
\end{abstract}

\section{Introduction}
 
The persistent rise in cardiovascular diseases (CVDs) is a leading cause of mortality worldwide, with an estimated 17.9 million fatalities attributed to CVDs in a year \cite{kauffman2020cardiovascular}. Early detection, timely intervention, and widely-accessible health-monitoring devices are crucial factors in preventing CVDs and reducing related healthcare costs. Democratizing accurate cardiac monitoring is critical, as it empowers individuals and communities to take control of their cardiovascular health regardless of socioeconomic status or geographic location \cite{bayoumy2021smart}. 

Electrocardiography (ECG) has long been the gold standard for cardiac monitoring, offering crucial insights to diagnose numerous CVDs \cite{hannun2019cardiologist}. However, their requirement for specialized equipment and skilled medical personnel limits their accessibility and affordability for daily life cardiac activity monitoring. Continuous ECG monitoring is always preferred over conventional clinical settings as it offers a comprehensive assessment of cardiac activity by collecting and analyzing data from patients uninterruptedly in their day-to-day life, and is not limited to a single point in time. For example, the detection of Atrial Fibrillation (AFib) remains challenging due to its asymptomatic and intermittent nature \cite{camm2012usefulness}, which often goes undetected in clinical settings that usually take place over the short duration of a patient's visit to the physician \cite{steinhubl2018effect}. 

To overcome these limitations, Photoplethysmography (PPG) has emerged as a promising non-invasive and cost-effective alternative for continuous cardiac monitoring in wrist-based wearable devices.
PPG is a non-invasive optical signal that measures blood volume fluctuations to capture cardiac activity. However, PPG lacks detailed cardiac information in comparison to ECG, which limits its use in diagnosing CVDs. Due to the non-invasive nature and ability of PPGs to be integrated into wearable devices, it provides an unmatched level of ubiquity. 

To leverage the convenience and continuous monitoring ability of PPG and the diagnostic utility of ECG, PPG-to-ECG translation has emerged as a promising, yet challenging alternative \cite{sarkar2021cardiogan,vo2021p2e,lan2023performer}. The goal of PPG-to-ECG translation is to use deep learning techniques to generate ECG signals from PPG, with the aim that intricate %
CVD-related information in PPG signals become more prominent and detectable in the output ECG.

We identify two key areas that require significant improvement in the context of PPG-to-ECG translation. 
\textbf{First}, recent advances in Diffusion models have shown great promise in producing hyper-realistic content with high fidelity \cite{saharia2022photorealistic, blattmann2023align, ruan2023mm}. Yet, to our knowledge, they have not been successfully explored for PPG-to-ECG translation. More specifically, designing Diffusion models capable of effectively learning the nuances of both spatial and temporal dynamics of bio-signals like ECG has not been explored.
\textbf{Second}, despite promising results in PPG-to-ECG translation, to our knowledge, experiments to comprehensively evaluate the predictive power of the generated ECG in detecting CVDs in comparison to the original PPG signals has not been conducted. 

In this paper, we leverage diffusion models \cite{sohl2015deep} and develop a framework for PPG-conditioned high-fidelity ECG generation. We observe that Diffusion models like Denoising Diffusion Probabilistic Models (DDPMs) \cite{ho2020denoising} struggle to capture the nuances of bio-signal translation due to their inability to identify regions of interest (ROIs), and therefore fail to learn both the global structures and fine-grained local details. The underlying reason for this limitation is rooted in the forward process, which uniformly adds noise to the entire signal without considering the morphological differences of different bio-signals (e.g., QRS complex in ECG or systolic peak in PPG). To overcome this challenge, we introduce a novel Diffusion model called Region-Disentangled Diffusion Model (RDDM) which is able to capture the complex temporal dynamics and region-specific intricate details of ECG signals. In particular, we introduce a novel forward process that selectively adds noise to specific ROIs in ECG and a reverse process that disentangles the denoising of QRS and Non-QRS regions. 
Our experiments show that our method outperforms state-of-the-art methods in PPG-to-ECG translation in terms of standard error metrics (e.g., RMSE). Moreover, to tackle the second problem mentioned earlier, we introduce CardioBench, a comprehensive evaluation benchmark for PPG-to-ECG translation. CardioBench consists of a series of $5$ cardiac-related tasks selected to evaluate the utility of the generated ECG signals. The tasks 
include \textbf{(1)} Heart rate (HR) estimation, \textbf{(2)} AFib detection, \textbf{(3)} Stress and Affect classification, \textbf{(4)} Diabetes detection, and \textbf{(5)} Blood Pressure (BP) estimation. Evaluating our proposed method on CardioBench demonstrates that the generated ECG signals can be effectively used for cardiac-related tasks, and can outperform the use of the original PPG for the same tasks by large margins.

\noindent Following is a summary of our contributions:
\begin{itemize}[noitemsep,nolistsep,leftmargin=*]

\item We introduce RDDM, a novel diffusion model for high-fidelity PPG-to-ECG translation. To the best of our knowledge, this is the first diffusion model for cross-modal signal-to-signal translation in the bio-signal domain. 

\item Our proposed RDDM introduces a novel approach that disentangles the diffusion process into two distinct components. The first component captures the global temporal structure of ECG signals, while the second component is dedicated to capture the fine-grained local details. This enables RDDM to generate high-fidelity ECG and do so in just $10$ sampling steps making RDDM highly-efficient.
 
\item To evaluate the quality and utility of the generated ECG, we introduce CardioBench, a comprehensive evaluation benchmark comprised of $5$ challenging cardiac-related tasks.
\item
Our extensive evaluations demonstrate that the ECG signals generated by RDDM are significantly better in quality than DDPMs and prior works in both standard quantitative measures and their performance on CardioBench tasks. 
\item 
We make the code public to the research community\footnote{https://github.com/DebadityaQU/RDDM}.

\end{itemize}

\section{Related work}

\textbf{ECG synthesis with Generative models.} 
Generative models, particularly Generative Adversarial Networks (GANs) have been used as a promising approach in ECG synthesis \cite{chen2022me, golany2019pgans}.
Recently, Diffusion models have also been used for ECG generation \cite{alcaraz2023diffusion,adib2023synthetic,neifar2023diffecg}. Unconditional ECG generation using 2D Gramian Angular Field representations has been explored in \cite{adib2023synthetic}. 
DDPMs have been used for ECG generation from clinical statements \cite{alcaraz2023diffusion}. DiffECG \cite{neifar2023diffecg} explored the application of DDPM for class-conditional ECG synthesis.

\textbf{PPG-to-ECG translation with Generative models.}
The research landscape of PPG-to-ECG translation remains relatively under-explored. Earlier attempts to solve this problem primarily focused on estimating various ECG parameters from PPG signals rather than direct reconstruction using methods like Feature selection \cite{banerjee2014photoecg}, Linear regression \cite{zhu2021learning}, and Dictionary learning \cite{tian2022cross}. Such methods suffer a degradation in performance when evaluating in a subject-independent manner.
CardioGAN \cite{sarkar2021cardiogan} made the first attempt at using Generative models for direct PPG-to-ECG translation. They used unpaired training using CycleGAN \cite{zhu2017unpaired} with an improved architecture by incorporating attention and dual time-frequency discriminators. Empirically, they demonstrated significantly better HR estimation from the generated ECG compared to the original PPG. Similarly, \cite{vo2021p2e} utilized conditional Wasserstein GANs for PPG-to-ECG translation. \cite{lan2023performer} developed a Transformer-based framework for reconstructing ECG from PPG, and further fine-tuned the model for CVD detection using original PPG and generated ECG together.

\section{Region-Disentangled Diffusion Model}

Diffusion models \cite{sohl2015deep} consist of a forward and a reverse process.
The forward diffusion process $q(\x_t|\x_{t-1})$
can be characterized as a Markov chain that progressively adds Gaussian noise at every timestep $t$:
\begin{align}
\label{eq:forward-ddpm}
\begin{split}q(\x_{T}|\x_0):=\prod_{t=1}^{T}q(\x_t|\x_{t-1}),\quad \mbox{where}\quad \\
    q(\x_t|\x_{t-1}):=\Nc(\x_t;\sqrt{1-\beta_t}\x_{t-1},\beta_t\Ib).
\end{split}
\end{align}
Here, $\x_0 \sim q(\x_0)$ is a clean signal and $\beta_t$ is a small positive constant obtained from a fixed variance schedule $\beta_1, \cdots, \beta_T$.
Let $\alpha_t := 1 - \beta_t$ and $\bar{\alpha_t} := \prod_{s=1}^t \alpha_s$. Then, the forward diffusion process yields a sample at timestep $t$, denoted by $\x_t$, which can be obtained in a single step as follows:
\begin{align}
\label{eq:ddpm}
\x_t=\sqrt{\bar{\alpha}_t}\x_0+\sqrt{1-\bar{\alpha}_t}\boldsymbol{\epsilon},\quad \mbox{where}\quad \boldsymbol{\epsilon}\sim\Nc(\textbf{0},\Ib).
\end{align}
Since the reverse step of the forward process, $q(\x_{t-1}|\x_t)$ is computationally infeasible, DDPMs \cite{ho2020denoising} maximize the variational lower bound (ELBO) using a parameterized Gaussian transition $p_\theta(\x_{t-1}|\x_t)$ with parameter $\theta$. Consequently, the reverse process is approximated as a Markov chain with a learned mean and a fixed variance, initiated from $p(\x_T) = \mathcal{N}(\x_T; \textbf{0}, \boldsymbol{I})$:
\begin{align}
p_\theta(\x_{0:T}):=p_\theta(\x_T)\prod_{t=1}^{T}p_\theta(\x_{t-1}|\x_{t}),\quad \\ \mbox{where,}\quad
p_\theta(\x_{t-1}|\x_{t}):=\Nc(\x_{t-1};{\boldsymbol\mu}_\theta(\x_t,t),\sigma_t^2\Ib).
\end{align}
\begin{align}\label{eq:mu}
\mbox{and,}\quad
{\boldsymbol\mu}_\theta(\x_t,t):=\frac{1}{\sqrt{\alpha}_t}\Big{(}\x_t-\frac{1-\alpha_t}{\sqrt{1-\bar{\alpha}_t}}\boldsymbol{\epsilon}_\theta(\x_t,t)\Big{)},
\end{align}
The diffusion model $\epsilon_{\theta}$ can be conditioned upon the input signal $c$, which makes the conditional objective:

\begin{align}
\label{eq:cond-objective}
 L(\theta):=\mathbb{E}_{t,\x_0,\boldsymbol{\epsilon}}\Big{[}|\boldsymbol{\epsilon}-\boldsymbol{\epsilon}_\theta(\sqrt{\bar{\alpha}_t}\x_0+\sqrt{1-\bar{\alpha}_t}\boldsymbol{\epsilon},c, t)|^2\Big{]}.
\end{align}

PPG-to-ECG translation can be performed by adopting Equation \ref{eq:cond-objective} using the PPG signal as input condition $c$ to generate the corresponding ECG signal $\x_0$. To enable the Diffusion model to learn both global structures and fine-grained local details of ECG for nuanced bio-signal translation, we propose RDDM which modifies the forward and reverse process of DDPMs as discussed below.

\begin{figure}[t]
    \centering
    \includegraphics[width=\columnwidth]{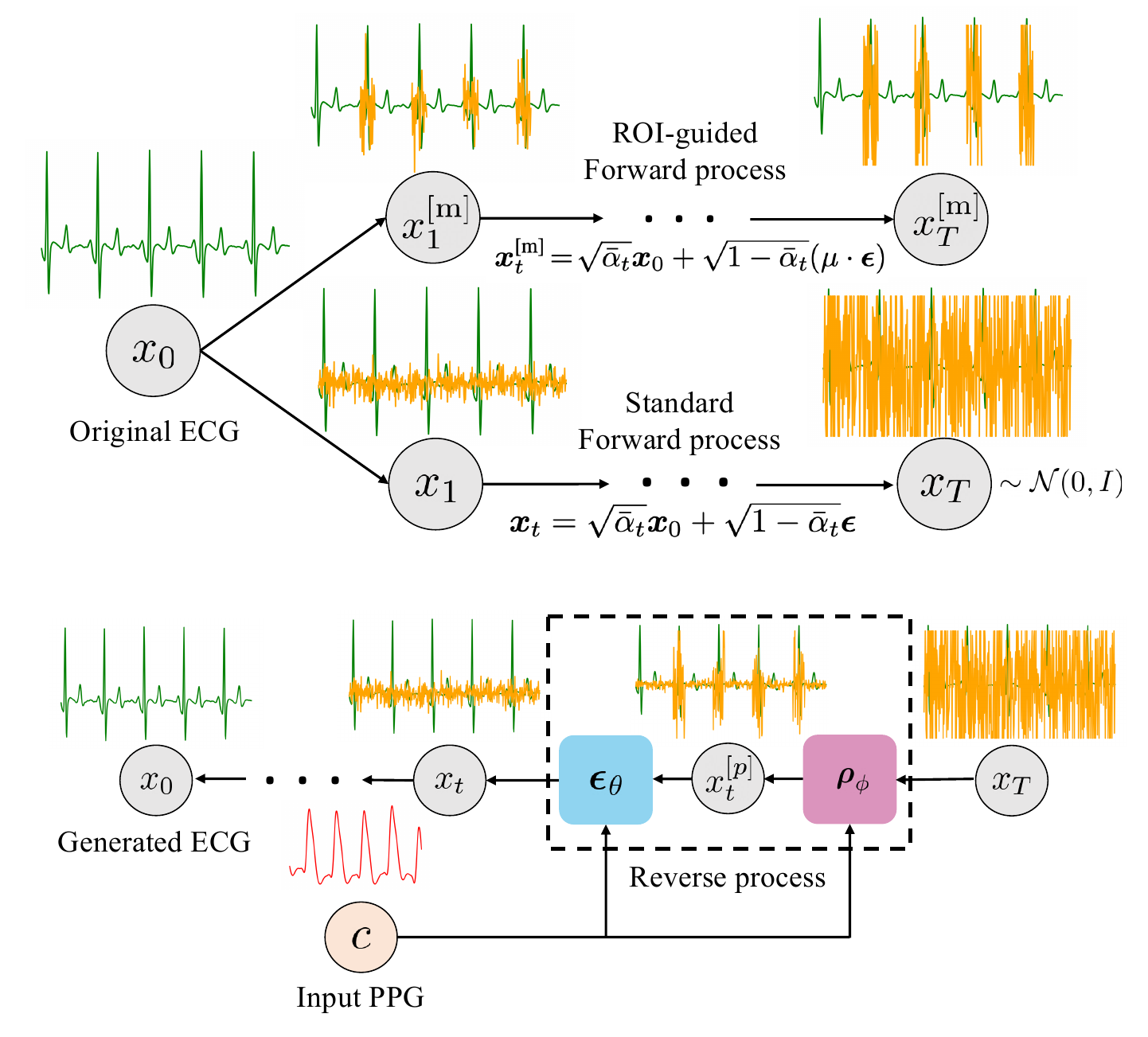}
    \caption{PPG-to-ECG translation with RDDM.  \textbf{Top:} ROI-guided Forward process selectively adds Gaussian noise to the QRS regions, while standard forward process adds noise uniformly across the ECG. \textbf{Bottom:} Our reverse process involves disentangled denoising of the QRS and Non-QRS regions by $\epsilon_{\theta}$ and $\rho_{\phi}$ respectively.
    }
    \label{fig:rddm}
\end{figure}

\noindent\textbf{Forward Process.} The forward process of RDDM consists of $2$ stages, (\textit{i}) we obtain a noise-added sample $\x_t$ following the standard forward process mentioned in Equation \ref{eq:ddpm}; (\textit{ii}) in parallel, we also obtain a selectively noise-added sample $\x_t^{[m]}$, in a process which we refer to as the ROI-guided forward process. Given the salience of cardiac-related information in the QRS complex of ECG signals, we define the ROI to enable the Diffusion model to learn these regions effectively alongside the global structure of the ECG which is facilitated through the standard forward process. Accordingly, we define a binary mask $\mu$ with the same length of $\x$ as:
\begin{equation}
\label{eq:rddm-roi}
\mu[i] =
\begin{cases}
1, & \text{if}\ i_r - \frac{\Gamma}{2} \leq i \leq i_r + \frac{\Gamma}{2} \\
0, & \text{otherwise}
\end{cases}
\end{equation} 
where, $i_r$ denotes the indices of R-peaks and $\Gamma$ is a hyper-parameter defined as the window size to capture the ROI centered at $i_r$.
Finally, the ROI-guided forward process is defined as: 
\begin{align}
\label{eq:rddm}
\x_t^{\text{[m]}}\!=\!\sqrt{\bar\alpha_t} \x_0 + \sqrt{1-\bar\alpha_t}(\mu \cdot \bepsilon) \mbox{ where }\boldsymbol{\epsilon}\sim\Nc(\textbf{0},\Ib).
\end{align} 

\noindent\textbf{Reverse Process.} Since $\mu$ is not available during inference, we approximate the parameterized transition $\rho_{\phi}(x_t^{\text{[m]}} | x_t)$ by our training objective defined as:
\begin{equation}
\begin{aligned}
\label{eq:rddm-objective}
 L(\theta, \phi):=\mathbb{E}\Big{[} \lambda_{1} | (\mu \cdot \boldsymbol{\epsilon})-\boldsymbol{\epsilon}_\theta(x_t^{\text{[m]}}, c,  t)|^2 \\ + \lambda_{2} |x_t^{\text{[m]}} - \rho_{\phi}(x_t, c, t)|^2\Big{]}.
\end{aligned}
\end{equation}
Here, $\lambda_{1}$ and $\lambda_{2}$ are the loss coefficients. In our proposed objective, the first part is responsible for learning to denoise the ROIs (QRS), while the second part learns to denoise the rest of the regions (Non-QRS), effectively disentangling the denoising of the important regions of ECG signals.

More specifically, the objective function first guides the learning of $\boldsymbol{\epsilon}_\theta$, where $\boldsymbol{\epsilon}_\theta$ is responsible for refining the ROIs. The loss term $|(\mu \cdot \boldsymbol{\epsilon})-\boldsymbol{\epsilon}_\theta(x_t^{\text{[m]}}, c, t)|^2$ encourages $\boldsymbol{\epsilon}_\theta$ to estimate noise that, when subtracted from the signal, refines and enhances the morphology of the ROIs. This is achieved by minimizing the difference between the noise generated by $\boldsymbol{\epsilon}_\theta$ and the ROI-specific noise $(\rho \cdot \boldsymbol{\epsilon})$ in the original data distribution. Therefore, \textbf{the role of $\boldsymbol{\epsilon}_\theta$ is to learn the fine-grained details of the ROI.}

\algrenewcommand\algorithmicindent{0.5em}
\begin{algorithm}[t]
  \caption{RDDM Training} \label{alg:training}
  \small
  \begin{algorithmic}[1]
    \Repeat
      \State $\bx_0 \sim q(\bx_0)$; 
      \State $t \sim \mathrm{Uniform}(\{1, \dotsc, T\})$
      \State $\bepsilon\sim\mathcal{N}(\bzero,\bI)$
      \State Apply ROI mask: $\bepsilon^{\text{[m]}} = (\mu \cdot \bepsilon)$
      \State Compute $x_t = \sqrt{\bar\alpha_t} \x_0 + \sqrt{1-\bar\alpha_t}\bepsilon$
      \State Estimate $x_t^{\text{[p]}} = \rho_{\phi}(x_t, c, t)$
      \State Compute $x_t^{\text{[m]}} = \sqrt{\bar\alpha_t} \x_0 + \sqrt{1-\bar\alpha_t}\bepsilon^{\text{[m]}}$
      
      \State Calculate gradient as
      \Statex $\grad_{\theta, \phi}\!\left( \lambda_{1} \left\| \bepsilon^{\text{[m]}}\!-\!\bepsilon_\theta(x_t^{\text{[m]}}, c, t) \right\|^2\! +\! \lambda_{2}  \left\| x_t^{\text{[m]}}\! -\! x_t^{\text{[p]}} \right\|^2 \right)$
    \Until{converged}
  \end{algorithmic}
\end{algorithm}

Second, the loss term $|x_t^{\text{[m]}} - \rho_{\phi}(x_t, c, t)|^2$ encourages $\rho_{\phi}$ to minimize the discrepancy between $x_t^{\text{[m]}}$ and the signal recovered by $\rho_{\phi}(x_t, c, t)$.
Consequently, the function $\rho_{\phi}$ learns to denoise the non-ROI parts of the ECG signal, which implicitly involves learning `where' to place the QRS complexes. 
\textbf{This results in $\rho_{\phi}$ capturing the temporal dynamics of the ECG signal}. This disentanglement makes RDDM particularly suitable for PPG-to-ECG translation, which requires capturing both the spatial and temporal characteristics of ECG signals. The overall training algorithm of RDDM has been outlined in Algorithm \ref{alg:training}. 

\begin{algorithm}[t]
  \caption{RDDM Sampling} \label{alg:sampling}
  \small
  \begin{algorithmic}[1]
    \State $\x_T \sim \mathcal{N}(\bzero, \bI)$
    \For{$t=T, \dotsc, 1$}
      \State $\bz \sim \mathcal{N}(\bzero, \bI)$ if $t > 1$, else $\bz = \bzero$
      \State $\x_t^{\text{[p]}} = \rho_{\phi}(x_t, c, t)$
      \State $\x_{t-1}\!=\!\frac{1}{\sqrt{\alpha_t}}\left( \x_t^{\text{[p]}} - \frac{1-\alpha_t}{\sqrt{1-\bar\alpha_t}} \bepsilon_\theta(\x_t^{\text{[p]}}, c, t) \right) + \sigma_t \bz$
    \EndFor
    \State \textbf{return} $\x_0$
  \end{algorithmic}
\end{algorithm}

\noindent\textbf{Sampling algorithm.} As outlined in Algorithm \ref{alg:sampling}, RDDM generates samples by initializing a random signal $\x_T$ from a Gaussian distribution and iteratively applying the reverse process. For each time step $t$ from $T$ to $1$, we sample a latent variable $\bz$ from a Gaussian distribution if $t > 1$, otherwise, $\bz = \bzero$. We then obtain the ROI masked signal $\x_t^{\text{[p]}}$ as $\rho_{\phi}(\x_t, c, t)$. Finally, we update the signal $\x_{t-1}$ using $\boldsymbol{\epsilon}_\theta$. The resulting signal $\x_0$ is the final output of the sampling algorithm. The forward and reverse processes of our proposed method are depicted in Figure \ref{fig:rddm}.

\section{Experiment setup}

\subsection{Implementation details}

\noindent\textbf{Datasets.} Following, we present the details of the datasets that are used in this study.

\begin{itemize}
    \item WESAD \citep{schmidt2018introducing} comprises approximately 24 hours of synchronized Lead-II ECG data (sampled at $700$Hz) and PPG data (sampled at $64$ Hz) from $15$ subjects with annotations of stress and affect states. %

    \item MIMIC AFib \cite{bashar2019noise} is a subset of the MIMIC-III waveform database \cite{johnson2016mimic} with binary labels for AFib. It consists of $35$ critically ill adults, out of which $19$ suffer from AFib. Both ECG and PPG signals are sampled at $125$ Hz. 

    \item PPG-BP \cite{liang2018new} consists of $657$ pre-windowed PPG signals from $219$ subjects, along with rich metadata describing the health conditions of each subject. The PPG was recorded at a sampling rate of $1$ kHz. Please note that unlike the other datasets used in this study, this dataset only consists of PPG signals and does not contain ECG.  %

    \item Cuffless BP \cite{kachuee2015cuff} is a dataset derived from the MIMIC-II waveform database \cite{saeed2011multiparameter}. It consists of thousands of PPG, ECG, and ABP signals recorded from various hospitals between 2001 and 2008, sampled at a frequency of $125$ Hz. %

    \item CAPNO \citep{karlen2021capnobase} comprises approximately $5.6$ hours of paired Lead-II ECG and PPG signals sampled at $300$ Hz. The data was recorded from $42$ subjects under medical supervision. 
    
    \item BIDMC \citep{pimentel2016toward} is a dataset collected from $53$ ICU patients, consisting of around $7$ hours of concurrent Multi-lead ECG (Lead II, V, and AVR) and PPG signals, both recorded at a frequency of $125$ Hz. We utilize only the Lead-II ECG in our work. 
    
    \item DALIA \citep{reiss2019deep} is a dataset with approximately $35$ hours of synchronized PPG and Lead-II ECG signals, recorded from $15$ subjects performing daily-life activities like walking and driving. The ECG and PPG were recorded at sampling rates of $700$ and $64$ Hz respectively. %

\end{itemize}

\noindent To train RDDM, we follow a similar setup to CardioGAN \cite{sarkar2021cardiogan} and create a diverse training set by combining the paired ECG and PPG signals from WESAD, DALIA, CAPNO, BIDMC, and MIMIC-AFib. In particular, we use the data from $80\%$ of the subjects for training and the remaining $20\%$ of subjects are used for cross-subject evaluation. Additionally, for CardioBench, we use WESAD, DALIA, MIMIC AFib, PPG-BP, and Cuffless BP.

\noindent \textbf{Data pre-processing.} 
We follow standard pre-processing steps for both ECG \cite{pan1985real, Makowski2021neurokit} and PPG \cite{Makowski2021neurokit}. First, we resample both signals at $128$ Hz. Subsequently, we apply a high-pass Butterworth filter with a cut-off frequency of $0.5$ Hz on ECG signals. Similarly, a band-pass Butterworth filter with a cut-off frequency between $0.5$ to $8$ Hz is applied on PPG signals. Additionally, subject-specific z-score normalization is applied to compensate for individual variances. Further, min-max scaling is applied to re-scale the ECG and PPG signals between $[-1, 1]$. Finally, we segment both signals into $4$-second windows which are used for RDDM training.

\noindent\textbf{Network architecture.} We adapt the UNet architecture from \cite{ho2020denoising, dhariwal2021diffusion} and modify for 1D signals, which is used as the backbone for $\epsilon_{\theta}$ and $\rho_{\phi}$ modules. Additionally, we replace the default self-attention blocks with cross-attention modules to better condition the generated ECG for a given PPG input \cite{saharia2022photorealistic}. We apply cross-attention between \textit{ECG at the previous timestep} and \textit{input PPG}. Diffusion timestep $t$ is added to each UNet block using Sinusoidal Position Embedding, similar to \cite{ho2020denoising, dhariwal2021diffusion, saharia2022photorealistic}.
The UNet architecture comprises $6$ down-sampling and $6$ up-sampling stages. During down-sampling, the number of filters is increased from $64$ to $1024$, doubling at each stage. Conversely, in up-sampling stages, the number of filters is decreased from $1024$ to $64$, halving at each stage. The final output is obtained from a single-channel convolutional layer. For a fair comparison, we also use the same architecture for the baseline DDPMs. 

\noindent \textbf{Training setup and hyper-parameters.} We train RDDM on $4\times$ NVIDIA A100 GPUs with a batch size of $512$. We choose AdamW \cite{loshchilov2018decoupled} as the optimizer with a Cosine learning rate scheduler, having a base learning rate of $10^{-4}$ for a total of $500$ epochs. 
We use a fixed linear variance scheduler with $\beta \in (0.0001, 0.2)$. We empirically set the sampling steps to $10$ and the ROI window size to $\Gamma = 32$ (see related study in Sec. \ref{sec:ablation}). The loss coefficients in Equation \ref{eq:rddm-objective} are set to $\lambda_{1} = 100$ and $\lambda_{2} = 1$. 

\begin{table*}[!ht]
    \centering
    \small
    \setlength{\tabcolsep}{2pt}
    \resizebox{1\linewidth}{!}{
    \begin{tabular}{@{}llllcccccllll@{}}
    \toprule
    \textbf{Task} & \textbf{Dataset(s)} & \textbf{Arch.} & \textbf{Input} & \textbf{Epochs} & \textbf{Optim} & \textbf{LR} & \textbf{Batch} & \textbf{Window} & \textbf{Metrics} & \textbf{GPUs}\\
    \toprule
    \textbf{HR} & \specialcellleft{DALIA \& \\WESAD} & - & ECG (raw) & - &  - & - & - & 8 sec & MAE & - \\ \midrule
    \textbf{AFib} & MIMIC-AFib & VGG-13 & ECG (STFT) & 25 & 
    \(\mathrm{AdamW}\) & \(1e^{-4}\)

 & 64 & 4 sec & F1, Acc. & $2\times$A100 \\ \midrule
    \textbf{Stress \& Affect} & WESAD & Transformer & ECG (raw) & 10 & 
    \(\mathrm{AdamW}\) & \(1e^{-4}\) & 64 & 60 sec & F1, Acc. & $2\times$A100\\ \midrule
    \textbf{Diabetes} & PPG-BP & VGG-11 & ECG (STFT) & 50 & \(\mathrm{AdamW}\) & \(1e^{-4}\) & 64 & 2 sec & F1, Acc. & $2\times$A100\\ \midrule
    \textbf{BP} & Cuffless BP & 1D UNet & ECG \& PPG (raw)  & 50 & \(\mathrm{Adam}\) & \(5e^{-4}\) & 512 & 10 sec &\specialcell{ MAE (SBP),\\MAE(DBP)} & $4\times$A100\\ 
    \bottomrule
    \end{tabular}
    }
    \caption{Implementation details for evaluation on CardioBench.}
    \label{appendix-table:implementation-details}
    \end{table*}

\subsection{Evaluation}

We perform two types of evaluation to ensure a robust assessment of generated ECGs. First, we perform `standard' evaluation to report the error/distance metrics between generated ECG and ground-truth ECG. Additionally, to understand the utility and predictive power of the generated ECG in analyzing cardiac activity, we introduce a comprehensive evaluation benchmark, CardioBench. Following, we provide the details for both setups.

\noindent \subsubsection{Standard evaluation.}
The generated ECG signals are evaluated on standard quantitative metrics in a subject-independent manner. Following CardioGAN \cite{sarkar2021cardiogan}, we generate 4-second ECG signals from RDDM and measure Root Mean Squared Error (RMSE) and Fréchet Distance (FD) with respect to the original ECG.

\subsubsection{CardioBench.}
To test the utility of the generated ECG in analyzing cardiac activity, e.g., detecting CVDs, we introduce CardioBench. CardioBench is designed as a comprehensive and challenging benchmark aimed at evaluating the generated ECG signals, with a focus on their ability to facilitate the detection of a range of heart conditions. Our proposed benchmark consists of $5$ cardiac-related tasks across $6$ datasets: 

    \textbf{(1) Heart rate estimation} on \textit{WESAD} \cite{schmidt2018introducing} and \textit{DALIA} \cite{reiss2019deep} datasets: Accurate HR estimation from synthetic ECG depends on the presence of R-peaks at the correct temporal locations with consistent RR intervals. Therefore, HR estimation is included to evaluate the temporal fidelity of generative models. In line with previous literature \cite{sarkar2021cardiogan}, we perform HR estimation by calculating Mean Absolute Error (MAE) in beats per minute (bpm) which measures the difference between the ground truth HR and the HR estimated from the generated ECG signals. We use \cite{hamilton2002open} for HR estimation from the generated ECG using $8$-second windows.
    
    \textbf{(2) Atrial Fibrillation detection} on \textit{MIMIC-AFib} \cite{bashar2019noise} dataset: 
    AFib detection in synthetic ECG demands accurate generation of non-periodic irregular rhythms, absent P waves, variable ventricular rates, possible fibrillatory waves, and temporally compressed QRS complexes \cite{burnsguidelines}. This ability serves as a vital measure of a generative model's proficiency in capturing both detailed morphological features and non-periodic temporal dynamics, crucial for AFib detection. We evaluate the binary AFib classification using $4$-second ECG windows. We apply  Short-time Fourier Transform (STFT) on the generated ECG and use it as input to a standard VGG-13 \cite{sallem2020detection}. 
    
    \textbf{(3) Stress and Affect classification} on \textit{WESAD} \cite{schmidt2018introducing} dataset: Stress and affect classification using generated ECG requires an accurate representation of nuanced changes in heart rate variability \cite{kim2018stress}, sympathetic-parasympathetic balance \cite{tewari2006sympathetic}, and emotional state-induced ECG patterns \cite{sarkar2020self}. This capability acts as an indicator of a generative model's adeptness at capturing subtle morphological features and dynamic temporal changes. We perform three-class classification of neutral, stress, and amusement. We adopt a Transformer-based classifier \cite{behinaein2021transformer} using a window size of $60$ seconds for a fair comparison with previous methods. 
    
    \textbf{(4) Diabetes detection} on \textit{PPG-BP} \cite{liang2018new} dataset: Diabetes detection from synthetic ECG involves accurate capture of the 
    anomalies in the ECG signals induced by the disease, such as prolonged QT intervals, alterations in the QRS complex, and changes in the ST-T segment \cite{veglio2004qt}. The ability to generate these distinguishing spatiotemporal variations serves as a crucial metric for ECG generation. We evaluate the generated ECG on the binary classification task of Diabetes detection.
    We adopt a VGG-11 classifier \cite{reiss2019deep} and apply STFT on the input. We use $2$-second windows as per the pre-windowed dataset. 
    
    \textbf{(5) Blood Pressure estimation} on \textit{Cuffless-BP} \cite{kachuee2015cuff} dataset: Estimation of BP from synthetic ECG require an accurate representation of ECG-derived features like heart rate, P wave, QRS duration, and QT interval, all of which correlate with blood pressure levels \cite{peng2006heart, bird2020assessment}. 
    This serves as a vital test of a generative model's capacity to reproduce fine-grained morphological and dynamic temporal features.
    We estimate the Systolic Blood Pressure (SBP) and Diastolic Blood Pressure (DBP), and evaluate using MAE.
    We use a simple 1D-UNet \cite{long2023bpnet} as our estimator, and use $10$ second windows following prior methods.

\noindent \textbf{Comparison setup.}
HR estimation using \cite{hamilton2002open}
does not require specific training, and generated ECG is used directly for estimation. The other $4$ tasks, however, require training neural nets. As a result, for AFib detection and Stress and Affect classification, we train the models using real ECG and evaluate on both real and generated ECG signals. We treat performance on the real ECG of the test splits as the upper bound and compare the performance of the generated ECG in each specific task. A similar approach is taken for BP estimation, with the distinction that it is treated as a multimodal task \cite{elgendi2019use}, for which we use the PPG and real ECG for training. Finally, to conduct Diabetes detection, we directly train the classifier on the generated ECG due to the unavailability of original ECG signals and use the ECG generated from the test set for testing. We present the key implementation details for evaluation on CardioBench in Table \ref{appendix-table:implementation-details}.

\section{Results}

\subsection{Standard evaluation}
The results presented in Table \ref{table:comparison} clearly show that RDDM surpasses the current state-of-the-art PPG-to-ECG translation model \cite{sarkar2021cardiogan}  by a significant margin on all the datasets. For instance, RDDM improves RMSE from $0.63$ to $0.24$ on BIDMC. Additionally, RDDM outperforms the best version of the baseline DDPM by a large margin even when RDDM uses just $10$ sampling steps compared to the best DDPM that uses $50$ sampling steps. 

\begin{table}[t]
\centering
\small
\begin{tabular}{@{}llcc@{}}
\toprule
\textbf{Dataset} & \textbf{Method} & \textbf{RMSE} $\mathbf{(\downarrow)}$ & \textbf{FD} $\mathbf{(\downarrow)}$ \\
\midrule 
\multirow{5}{*}{\textbf{WESAD}} & \cite{sarkar2021cardiogan} & 0.37 & 29.15\\
& DDPM (T = 10) & 0.36 & 30.58 \\
& DDPM (T = 25) & 0.29 & 14.50 \\
& DDPM (T = 50) & 0.26 & 9.17 \\ 
\cmidrule{2-4}
& \textbf{RDDM (ours)} & \textbf{0.21} & \textbf{3.93}\\
\midrule
\multirow{5}{*}{\textbf{CAPNO}} & \cite{sarkar2021cardiogan} & 0.38 & 31.10\\
& DDPM (T = 10) & 0.33 & 21.56\\
& DDPM (T = 25) & 0.28 & 8.98\\
& DDPM (T = 50) & 0.24 & 5.15 \\
\cmidrule{2-4}
& \textbf{RDDM (ours)} & \textbf{0.19} & \textbf{2.94}\\
\midrule
\multirow{5}{*}{\textbf{DALIA}} & \cite{sarkar2021cardiogan} & 0.42 &  27.51\\
& DDPM (T = 10) & 0.38 & 19.58\\
& DDPM (T = 25) & 0.32 & 7.62\\
& DDPM (T = 50) & 0.30 & 5.95 \\
\cmidrule{2-4}
& \textbf{RDDM (ours)} & \textbf{0.25} & \textbf{5.84}\\ 
\midrule
\multirow{5}{*}{\textbf{BIDMC}} & \cite{sarkar2021cardiogan} & 0.63 & 154.64\\
& DDPM (T = 10) & 0.37 & 22.00 \\
& DDPM (T = 25) & 0.33 & 10.38 \\
& DDPM (T = 50) & 0.29 & 7.35\\
\cmidrule{2-4}
& \textbf{RDDM (ours)} & \textbf{0.24} & \textbf{6.72}\\
\midrule
\multirow{5}{*}{\textbf{MIMIC}} & \cite{sarkar2021cardiogan} & 0.54 & 99.03\\
& DDPM (T = 10) & 0.36 & 20.97 \\
& DDPM (T = 25) & 0.29 &  8.75 \\
& DDPM (T = 50) & 0.27 & 6.72\\
\cmidrule{2-4}
& \textbf{RDDM (ours)} & \textbf{0.22} & \textbf{6.71}\\
\bottomrule
\end{tabular}
\caption{Quantitative comparison with the state-of-the-art methods for \textbf{PPG-to-ECG translation}. RDDM consistently outperforms prior works by a large margin in all the setups. Moreover, RDDM also outperforms the best DDPM which uses $50$ sampling steps compared to ours using $10$. 
}
\label{table:comparison}
\end{table}

\subsection{Evaluation on CardioBench}

\noindent\textbf{HR estimation. } Table \ref{table:heart-rate} demonstrates that RDDM outperforms prior works and DDPM by achieving a substantially lower MAE on both datasets. On DALIA, RDDM achieves state-of-the-art MAE of $4.49$ bpm, outperforming the best PPG-based method \cite{song2021ppg} by $1.53$ bpm. RDDM demonstrates an improvement of $1.17$ bpm over DDPM. Next, on WESAD, RDDM achieves state-of-the-art MAE of $1.40$ bpm compared to the best PPG-based method \cite{reiss2019deep}, CardioGAN, and DDPM, which achieve $9.50$, $8.60$, and $3.22$ respectively.

\begin{table}[t]
    \centering
    \small
    \setlength{\tabcolsep}{3pt}
    \resizebox{1\linewidth}{!}{
\begin{tabular}{@{}lllc@{}}
    \toprule
    \textbf{Dataset} & \textbf{Method} & \textbf{Test modality} & \textbf{MAE}
    $(\downarrow)$\\
    \toprule
    \multirow{8}{*}{\textbf{DALIA}} & \cite{schack2017computationally} & Orig. PPG & 20.5\\
    & \cite{reiss2019deep} & Orig. PPG & 11.1 \\
    & \cite{wojcikowski2020photoplethysmographic} & Orig. PPG & 6.77\\
    & \cite{song2021ppg} & Orig. PPG & 6.02\\
    & \cite{sarkar2021cardiogan} & Gen. ECG &  8.30\\
    & DDPM (T = 50) & Gen. ECG & 5.66 \\
    \cmidrule{2-4}
    & \textbf{RDDM (ours)} & \textbf{Gen. ECG} & \textbf{4.49}\\
    \midrule
    \multirow{6}{*}{\textbf{WESAD}} & \cite{schack2017computationally} & Orig. PPG & 19.90\\
    & \cite{reiss2019deep} & Orig. PPG & 9.50\\
    & \cite{sarkar2021cardiogan} & Gen. ECG & 8.60\\
    & DDPM (T = 50) & Gen. ECG & 4.67 \\
    \cmidrule{2-4}
    & \textbf{RDDM (ours)} & \textbf{Gen. ECG} & \textbf{1.40}\\
    \bottomrule
    \end{tabular}
    }
    \caption{Evaluation of \textbf{Heart rate estimation.} RDDM outperforms the prior works by a significant margin.} 
    \label{table:heart-rate}
    \end{table}
    
\noindent\textbf{AFib detection. } 
As reported in Table \ref{table:afib}, RDDM outperforms both existing PPG-based methods and DDPM. Specifically, RDDM achieves F1 $\!=\!0.71$ and Acc. $= 0.65$, demonstrating improvements of $11\%$ and $22\%$ over DDPM, and approximately $16\%$ and $4\%$ when compared to the top-performing PPG-based method \cite{aliamiri2018deep}. Moreover, our method shows the closest performance to the upper bound, achieving an F1 of $0.71$ vs. the upper bound $0.74$. 
    \begin{table}[t]
    \centering
    \small
    \setlength{\tabcolsep}{3pt}
    \resizebox{0.99\linewidth}{!}{
    \begin{tabular}{@{}lllcc@{}}
    \toprule
    \textbf{Test modality} & \textbf{Classifier} & \textbf{Acc.}$(\uparrow)$ & \textbf{F1}$(\uparrow)$ \\
    \toprule
    Orig. ECG (Upper bound) & VGG-13 & 0.73 & 0.74 \\ \midrule
    Orig. PPG & \cite{aliamiri2018deep}$^*$ & 0.61 & 0.55\\
    Orig. PPG & \cite{shen2019ambulatory}$^*$ & 0.51 & 0.47 \\
    Gen. ECG (DDPM (T = 50)) & VGG-13 & 0.43 & 0.60\\
    \cmidrule{1-4}
    \textbf{Gen. ECG (RDDM (ours))} & \textbf{VGG-13} & \textbf{0.65} &  \textbf{0.71}\\

    \bottomrule
    \end{tabular}
    }
    \caption{Evaluation of \textbf{AFib detection} on MIMIC subset. RDDM significantly outperforms prior works and reports an F1 score close to the upper bound.
    $^*$re-implementation by us.}
    \label{table:afib}
    \end{table}

    \begin{figure*}[t]
    \centering
    \includegraphics[width=0.85\linewidth]{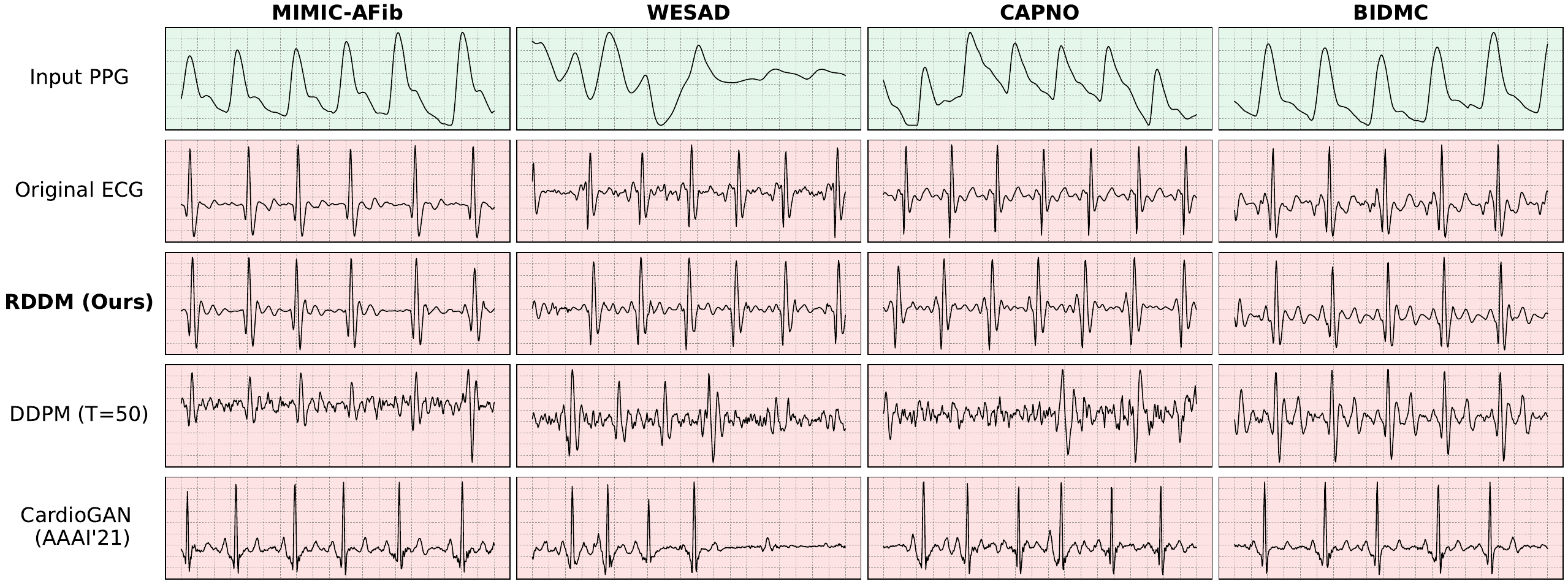}
    \caption{
    \textbf{Qualitative comparison} on generated ECG using varying nature of input PPG. RDDM consistently generates ECG that closely resembles the original ECG, even when the input PPG is noisy (e.g., WESAD), unlike CardioGAN and DDPM.
    }
    \label{fig:qualitative}
    \end{figure*} 
    
\noindent\textbf{Stress and Affect classification.} As highlighted in Table \ref{table:wesad}, our method achieves superior performance compared to prior methods, with F1 $= 0.64$ and Acc. $= 0.71$. Compared to the best PPG-based method \cite{schmidt2018introducing}, RDDM exhibits a boost of $7\%$ in F1 and $6\%$ in Acc. Furthermore, RDDM also outperforms DDPM which achieves similar performance to \cite{schmidt2018introducing}. Moreover, our method shows the closest performance to the upper bound, achieving an F1 of $0.64$, compared to the upper bound F1 of $0.68$.

    \begin{table}[t]
    \centering
    \small
    \setlength{\tabcolsep}{3pt}
    \resizebox{1\linewidth}{!}{
    \begin{tabular}{@{}llcc@{}}
    \toprule
    \textbf{Test modality} & \textbf{Classifier} & \textbf{Acc.}$(\uparrow)$ & \textbf{F1}$(\uparrow)$\\
    \toprule
    Orig. ECG (Upper bound) & \cite{behinaein2021transformer}  & 0.75 & 0.68\\ \midrule
    Orig. PPG &  \cite{schmidt2018introducing} & 0.65 & 0.57 \\ 
    Orig. PPG &  \cite{lisowska2021catching} & 0.58 & - \\
    Gen. ECG (DDPM (T = 50)) & \cite{behinaein2021transformer} & 0.65 & 0.58\\ \midrule
    \textbf{Gen. ECG (RDDM (ours))} & \textbf{\cite{behinaein2021transformer}} & \textbf{0.71} & \textbf{0.64}\\
    \bottomrule
    \end{tabular}
    }
    \caption{Evaluation of \textbf{Stress and Affect detection} on WESAD dataset. RDDM not only outperforms prior works, also achieves a very close performance to the upper bound.}
    \label{table:wesad}
    \end{table}
     
\noindent\textbf{Diabetes detection.} As illustrated in Table \ref{tab:diabetes}, RDDM significantly improves upon existing PPG-based methods and DDPM in diabetes detection. Notably, our method achieves an Acc. of $0.80$ and an F1 of $0.52$, indicating substantial improvements of $15\%$ in Acc. and $8\%$ in F1, compared to the leading PPG-based method \cite{avram2020digital}, which registers an Acc. of $0.65$ and an F1 of $0.44$. Furthermore, RDDM outperforms DDPM by an impressive margin of $33\%$ and $9\%$ in terms of Acc. and F1 respectively. It is notable that despite the multimodal classifier introduced by Performer \cite{lan2023performer} being pre-trained on a large corpus, a classifier trained from scratch only on ECG signals generated by RDDM still outperforms it, marking an Acc. improvement of $4\%$. This result shows the high-fidelity of the ECG generated by RDDM.

    \begin{table}[t]
    \centering
    \small
    \setlength{\tabcolsep}{3pt}
    \resizebox{1\linewidth}{!}{
    \begin{tabular}{@{}llcc@{}}
    \toprule
    \textbf{Test modality} & \textbf{Classifier}  & \textbf{Acc.}$(\uparrow)$ & \textbf{F1}$(\uparrow)$\\
    \toprule
    Orig. PPG & \cite{avram2020digital}$^*$ &  0.65 & 0.44\\ 
    Gen. ECG (Performer) + Orig. PPG & \cite{lan2023performer}$^{**}$  & 0.76 & - \\  
    Gen. ECG (DDPM (T=50)) & VGG-11 &  0.47 & 0.43\\ \midrule
    \textbf{Gen. ECG (RDDM (ours))} & \textbf{VGG-11} & \textbf{0.80} & \textbf{0.52}\\ 
    \bottomrule
    \end{tabular}
    }
    \caption{Evaluation of \textbf{Diabetes detection} on PPG-BP dataset. 
    RDDM outperforms both uni-modal (Orig. PPG) and multimodal (Gen. ECG + Orig. PPG) prior works. 
    $^{*}$re-implementation by us. $^{**}$uses pre-trained weights.}
    \label{tab:diabetes}
    \end{table}

    \begin{figure*}[]
    \centering

    \includegraphics[width=\textwidth]{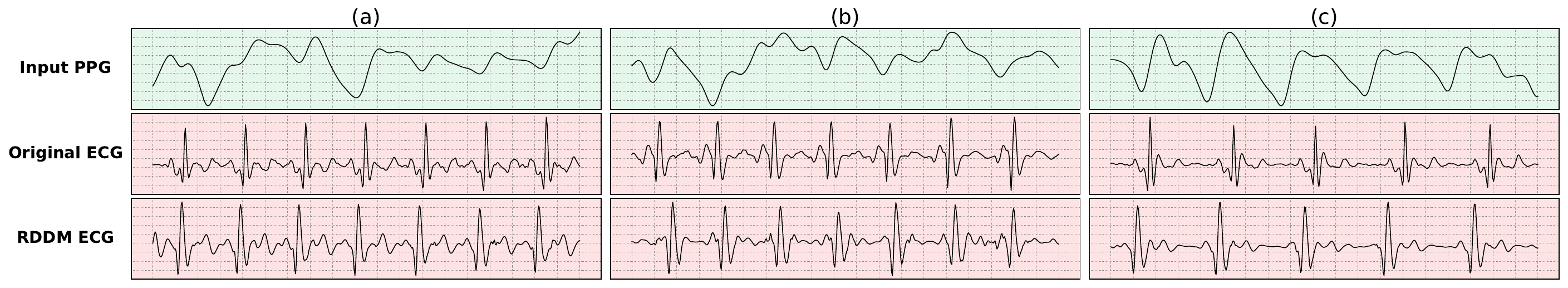}
    \caption{
    RDDM-generated ECG signals of subjects from the WESAD dataset experiencing different affective states. 
    }
    \label{fig:supp-stress}

    \centering
    \includegraphics[width=\textwidth]{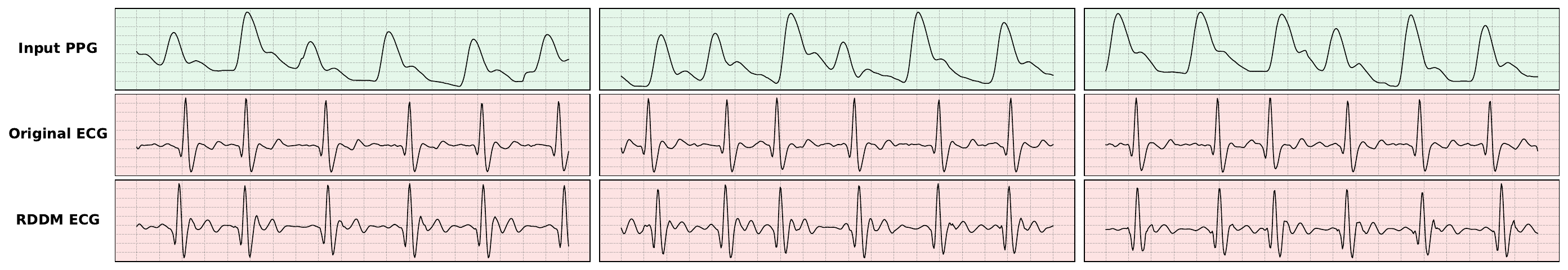}
    \caption{RDDM-generated ECG signals of $3$ AFib patients from the test set of MIMIC-AFib dataset.}
    \label{fig:supp-afib}
\end{figure*}

\noindent\textbf{BP estimation.}  As demonstrated in Table \ref{table:bp}, RDDM attains a state-of-the-art performance in BP estimation, achieving an MAE-SBP of $2.48$ and an MAE-DBP of $1.20$. Remarkably, RDDM makes significant improvements compared to the best PPG-based method \cite{vardhan2021bp} with a margin of $2.68$ in MAE-SBP and $2.25$ in MAE-DBP. This performance gain supports the notion that the ECG generated by RDDM provides complementary information learned through training on a large dataset of PPG-ECG pairs, which could enhance the performance beyond what is possible with PPG-only BP estimation methods. Furthermore, RDDM even outperforms the upper bound by achieving MAE-SBP $= 2.48$ vs. the upper bound MAE-SBP $= 3.31$, and performs closely to the upper bound in terms of MAE-DBP. We believe this improvement over upper-bound is observed as the original ECG signals were quite noisy. Although DDPM achieves the lowest MAE-DBP $= 0.63$, it performs poorly in terms of estimating SBP, with an MAE-SBP $=5.95$. In contrast, RDDM provides a balanced performance for estimating both SBP and DBP, demonstrating real-world utility.

    \begin{table}[t]
    \centering
    \small
    \setlength{\tabcolsep}{2pt}
    \resizebox{1\linewidth}{!}{
    \begin{tabular}{@{}llcc@{}}
    \toprule
    \textbf{Test modality} & \textbf{Estimator} & \textbf{MAE-SBP}$(\downarrow)$ & \textbf{MAE-DBP}$(\downarrow)$\\
    \toprule
    \specialcellleft{Orig. PPG\\~~+ Orig. ECG (Upper bound)} & UNet & 3.31 & 0.75\\
    \midrule
    Orig. PPG & \cite{ibtehaz2022ppg2abp} & 5.73 & 3.45\\
    Orig. PPG & \cite{vardhan2021bp} & 5.16 & 2.89\\
    \specialcellleft{Orig. PPG\\ ~~+ Gen. ECG (DDPM (T = 50))} & UNet & 5.95 & \textbf{0.63}\\
    \midrule
    \specialcellleft{\bf Orig. PPG \\\bf ~~+ Gen. ECG (RDDM (ours))} & \textbf{UNet} & \textbf{2.48} & 1.20\\
    \bottomrule
    \end{tabular}
    }
    \caption{Evaluation of \textbf{BP estimation} on Cuffless-BP dataset. RDDM clearly outperforms prior methods, and even shows a lower MAE-SBP than the upper bound.} 
    \label{table:bp}
    \end{table}

\begin{figure*}[!h]
    \centering
    \setlength\tabcolsep{0pt}
    \begin{tabular}{ccccc}
        \includegraphics[width=0.2\linewidth]{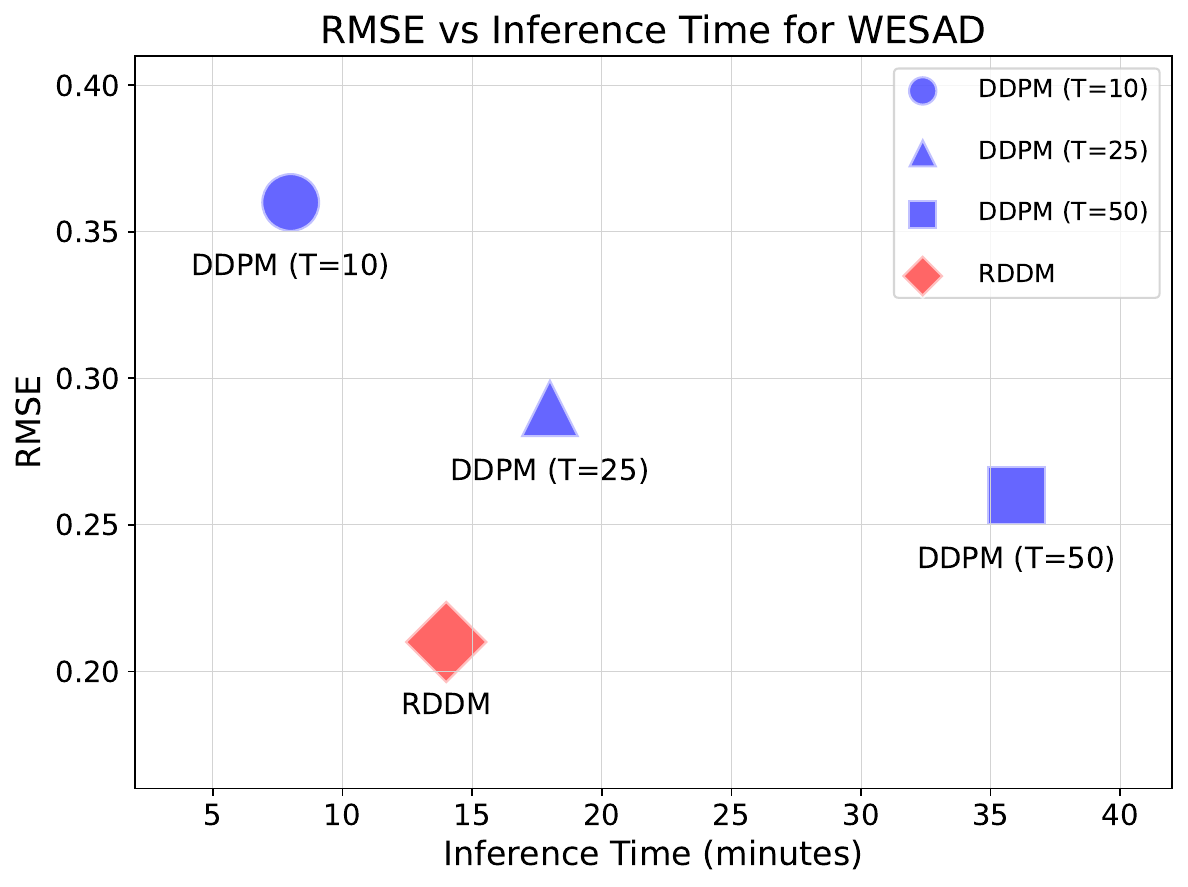} &
        \includegraphics[width=0.2\linewidth]{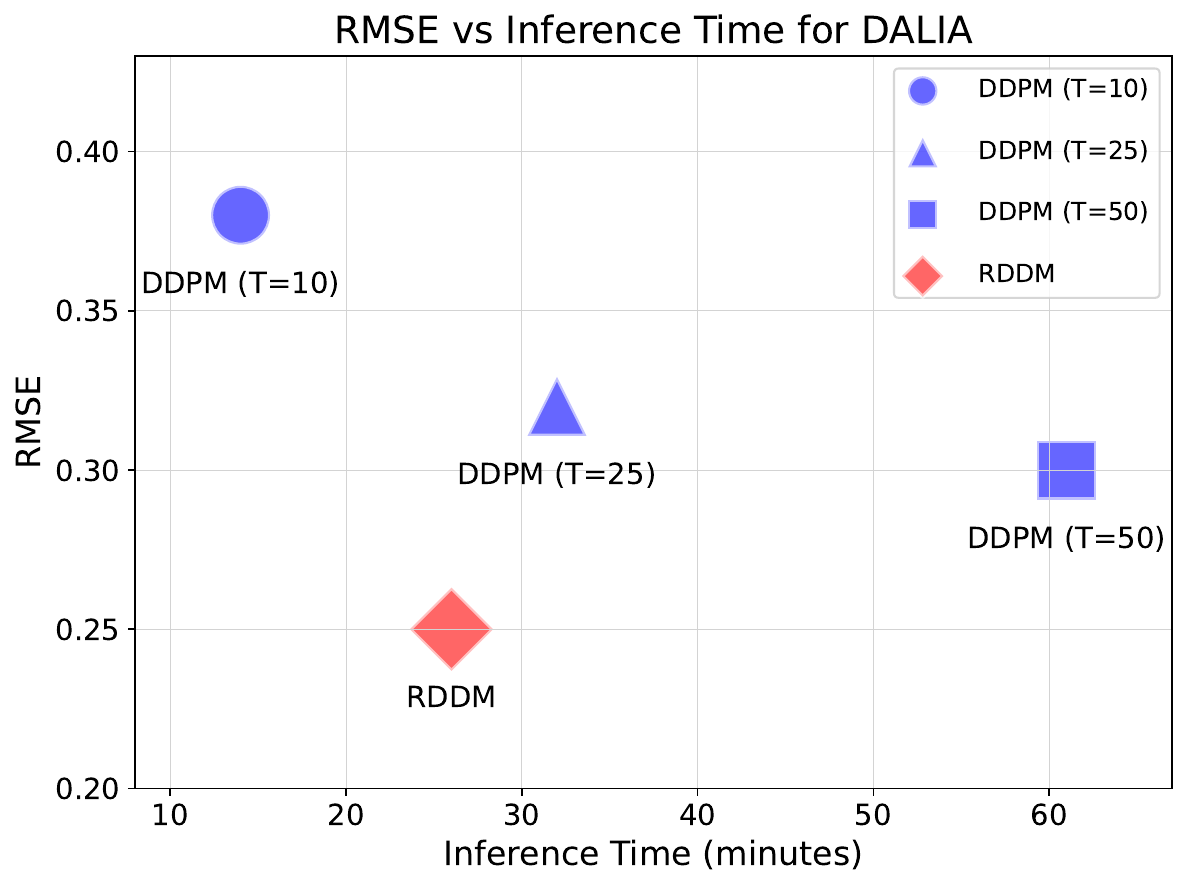} &
        \includegraphics[width=0.2\linewidth]{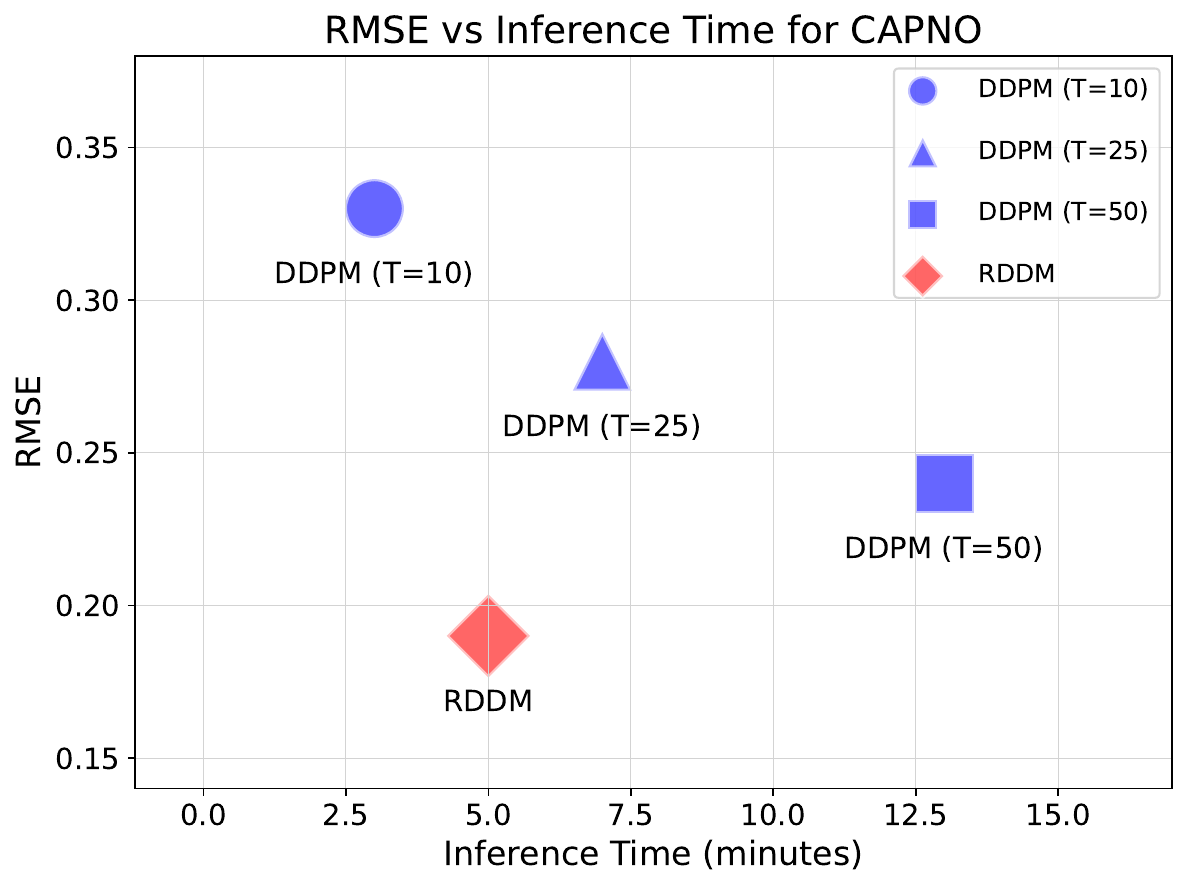} &
        \includegraphics[width=0.2\linewidth]{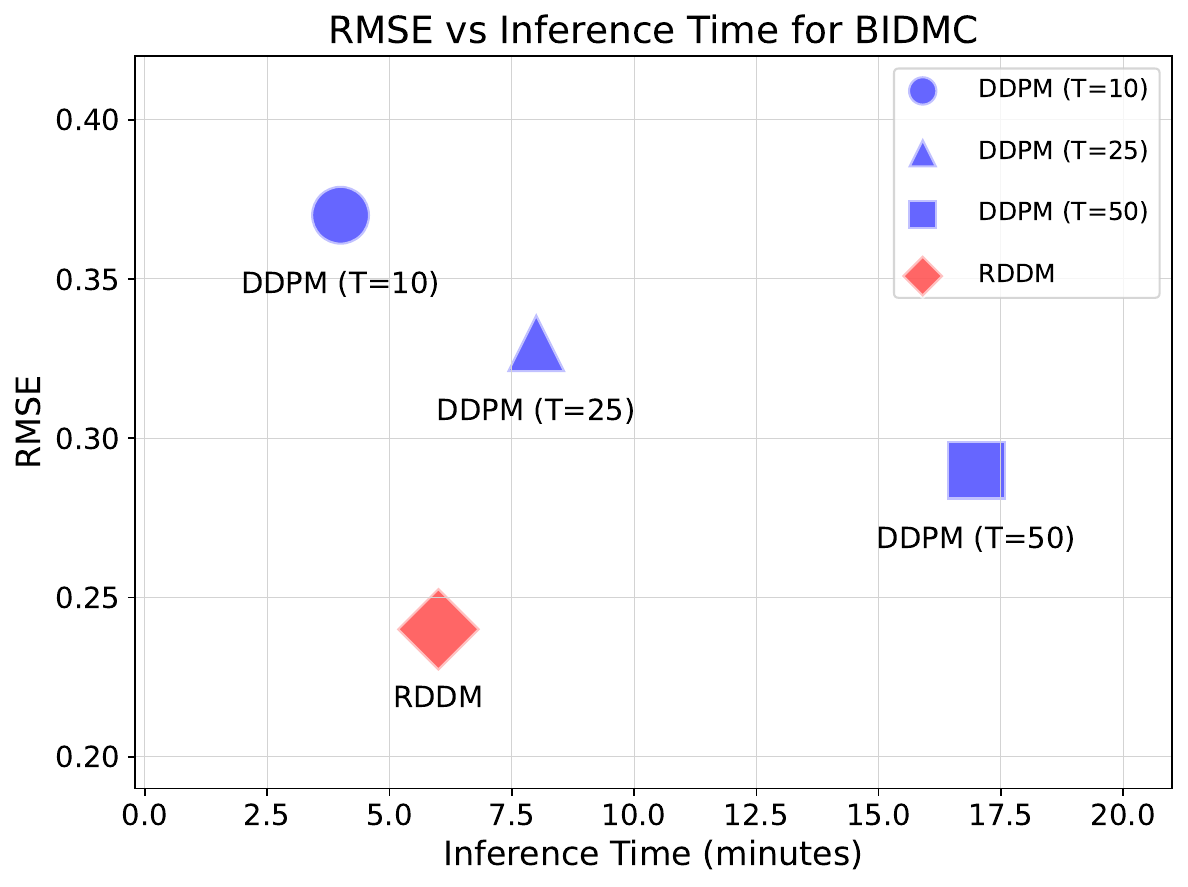} &
        \includegraphics[width=0.21\linewidth]{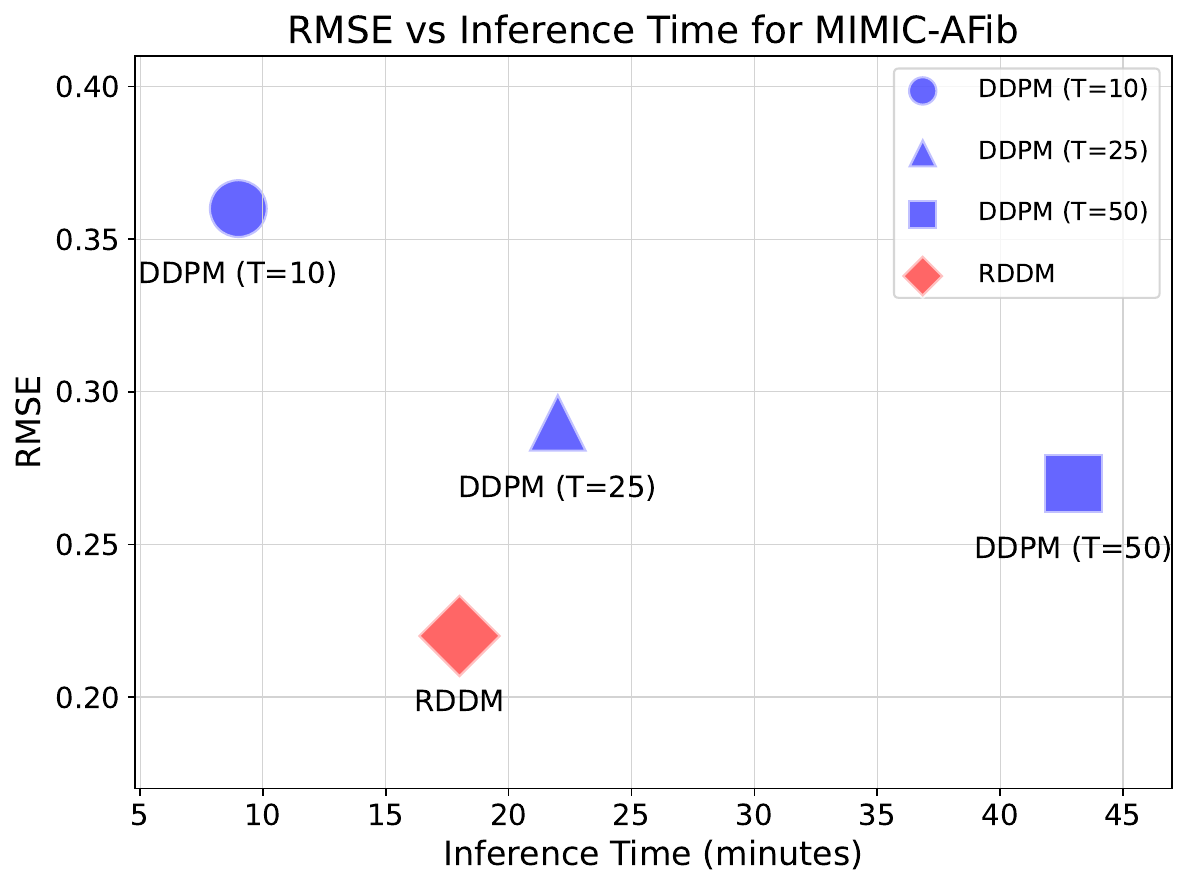} \\
    \end{tabular}
    \caption{\textbf{Inference Time Comparison:} 
    The plot illustrates the tradeoff between RMSE and inference time. RDDM demonstrates the best balance across all datasets, outperforming DDPM with (T$=50$) even though the latter is approx. $2.5 \times$ slower, highlighting the efficiency and effectiveness of the proposed method.
    }

    \label{fig:inference-time}
\end{figure*}

\subsection{Qualitative analysis}
\label{section:qualitative-analysis}
In Figure \ref{fig:qualitative}, we perform qualitative comparisons between RDDM with the best DDPM ($50$ sampling steps) and CardioGAN. Our comparison includes samples from MIMIC-AFib, WESAD, CAPNO, and BIDMC, covering a diverse set of PPG inputs of varying natures. The ECG signals generated by RDDM demonstrate high-fidelity compared to the ground-truth ECG signals, across all datasets. While comparing samples from MIMIC and BIDMC, generated ECG signals from CardioGAN and DDPM fail to accurately reproduce the QRS complex structure and significantly deviate from the original ECG morphology. Additionally, the generated ECG samples from DDPM exhibit higher noise as shown in MIMIC. In stark contrast, RDDM generates ECG signals that are almost identical to the ground truth ECG. Furthermore, in case of noisy PPG input, RDDM maintains temporal fidelity (as shown in WESAD) and also approximates the fine-grained structure of the QRS complex with remarkable precision. In contrast, DDPM and CardioGAN struggle to accurately generate QRS complex, leading to distorted temporal fidelity which results in irregular RR intervals. We provide additional generated ECG samples corresponding to different affective states in Figure \ref{fig:supp-stress} and samples corresponding to AFib patients in Figure \ref{fig:supp-afib}.

\subsection{Inference Time Comparison} 
Figure \ref{fig:inference-time} presents an inference time comparison between RDDM and DDPM variants. We measure the total inference time required to translate the PPG test set of each dataset into ECG signals for standard evaluation (i.e., RMSE). We use a single Nvidia 2080Ti GPU for this experiment. Across all datasets, RDDM consistently exhibits the best balance between RMSE and inference time. In contrast, DDPM with $50$ sampling steps not only performs worse in RMSE but is also slower by a factor of approx. $2.5 \times$, highlighting the superiority of RDDM in both speed and fidelity.

\subsection{Analysis of RDDM}
\label{sec:ablation}

\begin{figure}[t]
    \centering
    \setlength\tabcolsep{0pt}
    \begin{tabular}{cc}
    \includegraphics[width=0.45\linewidth,height=0.48\linewidth]{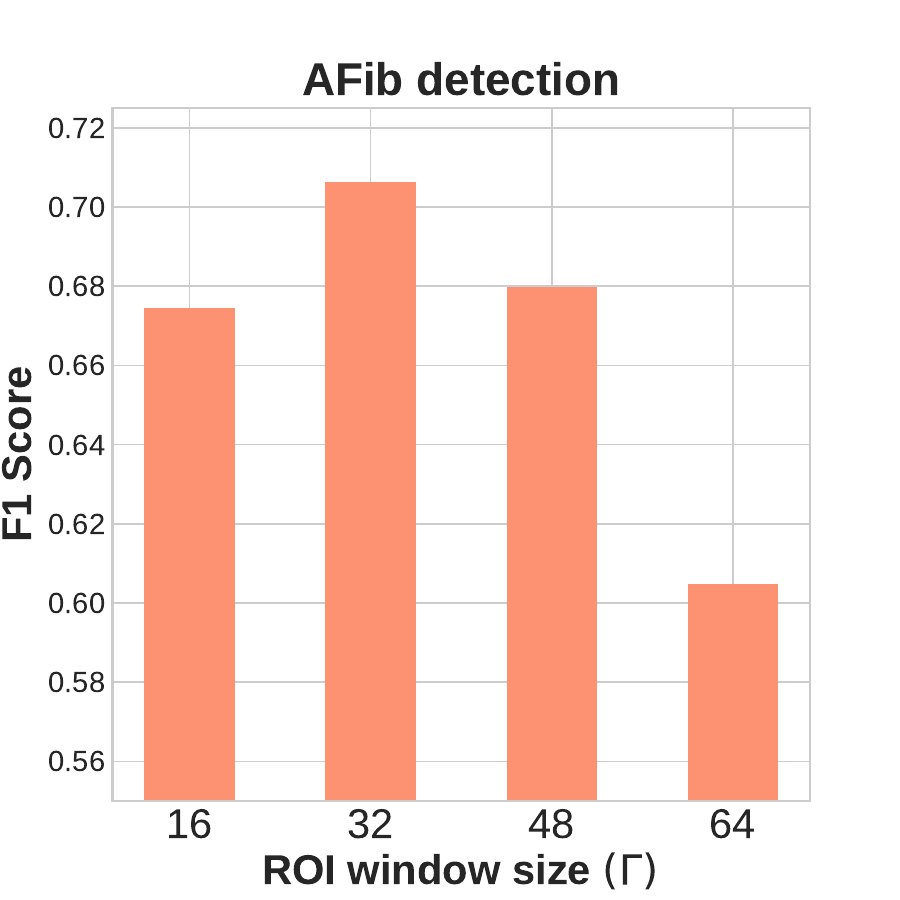}
         &  
    \includegraphics[width=0.45\linewidth]{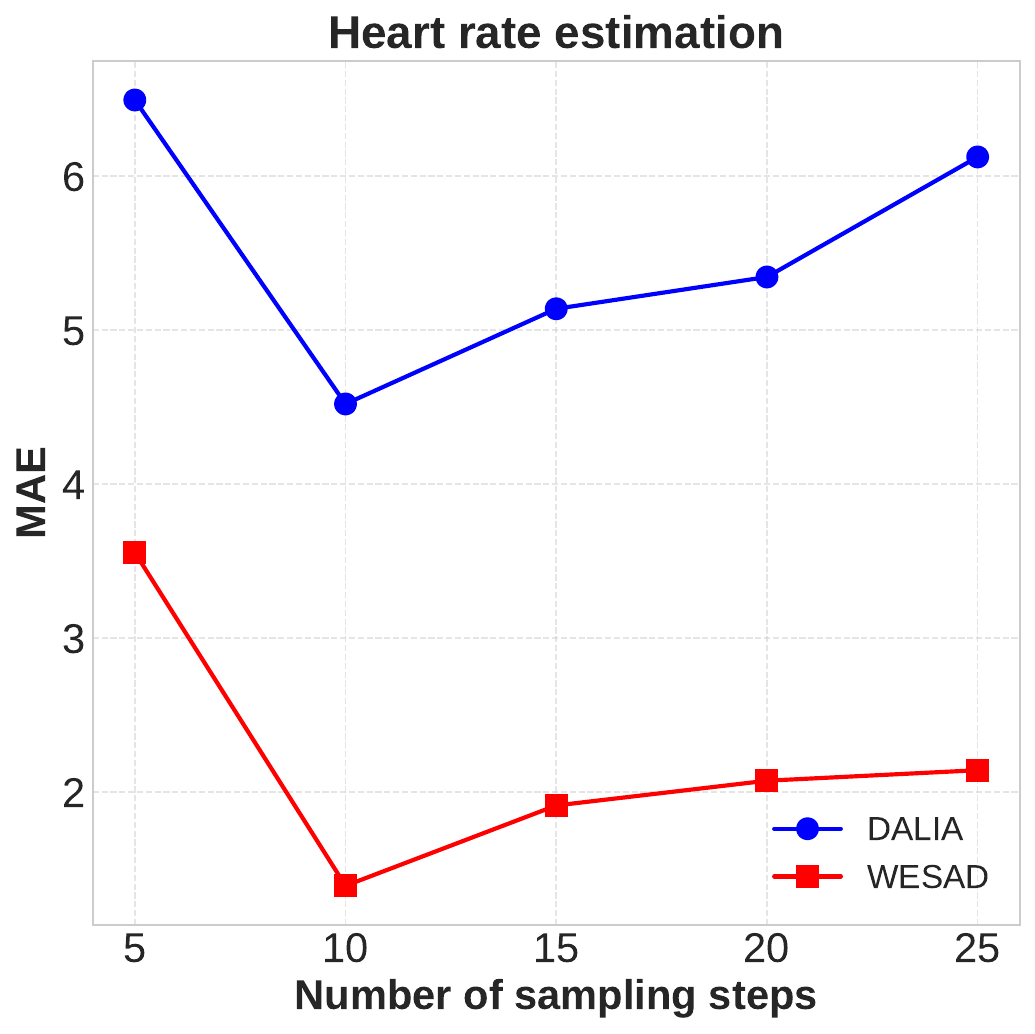}
    \\
    \end{tabular}
    \caption{
    \textbf{Left:} RDDM shows an improvement in performance when $\Gamma$ is increased from $16$ to $32$, but beyond that, the performance degrades.
    \textbf{Right:} On both datasets, RDDM achieves the best performance in just $10$ sampling steps.
    }
    \label{fig:ablation}
\end{figure}

\textbf{ROI window size.} 
To find the optimal ROI window size, we vary the value of $\Gamma$ between $16$ and $64$ with a step size of $16$ and track their performance on AFib detection. Figure \ref{fig:ablation} (left) shows that both extremely small ($\Gamma\!=\!16$) and large window ($\Gamma\!=\!64$) sizes degrade performance as they fail to capture the ROI accurately. An ROI window size of $32$ works best in our setup, which is used in all the experiments.

\noindent \textbf{Sampling steps.} To find the optimal value, we vary the number of sampling steps from $5$ to $25$ with a step size of $5$ and evaluate on HR estimation. 
The results presented in Figure \ref{fig:ablation} (right) exhibit a similar trend for both WESAD and DALIA. The performance significantly improves when the number of sampling steps is increased from $5$ to $10$. However, the performance declines when the sampling steps are increased further. Thus, we set the sampling steps as $10$ for RDDM.

\section{Conclusion}
In this paper, we introduce a novel diffusion model RDDM, specifically designed for high-fidelity PPG-to-ECG translation. To the best of our knowledge, RDDM is the first diffusion model for cross-modal signal-to-signal translation within the domain of bio-signals. The issue of indiscriminate noise addition during the forward process limits diffusion models like DDPMs from capturing both temporal structures and morphological details. RDDM improves upon them by introducing a novel ROI-guided forward process and by disentangling the reverse process into $2$ distinct components: (\textit{i}) capturing the global temporal structure of ECG signals and (\textit{ii}) capturing fine-grained local details. This allows RDDM to generate high-fidelity ECG signals in just $10$ sampling steps. To assess the utility of generated ECG, we introduce a comprehensive evaluation benchmark CardioBench.
CardioBench includes $5$ popular cardiac-related tasks: HR and BP estimation, stress classification, and detection of AFib and diabetes. RDDM achieves superior performance compared to DDPM and prior works, excelling in both standard quantitative measures and performance on CardioBench.

\section*{Acknowledgement}
This work was supported by Mitacs, Vector Institute, and Ingenuity Labs Research Institute.

\bibliography{main}

\end{document}